# Pneumatic capillary gun for ballistic delivery of microparticles

Dmitry Rinberg

*Monell Chemical Senses Center, 3500 Market St., Philadelphia, PA 19104*

Claire Simonnet and Alex Groisman[a)]

*Department of Physics, University of California, San Diego, 9500 Gilman Drive, La Jolla, CA, 92093*

**Abstract**

A pneumatic gun for ballistic delivery of microparticles to soft targets is proposed and demonstrated. The particles are accelerated by a high speed flow of Helium in a capillary tube. Vacuum suction applied to a concentric, larger diameter tube is used to divert substantially all of the flow of Helium from the gun nozzle, thereby preventing the gas from hitting and damaging the target. Speed of ejection of micron-sized gold particles from the gun nozzle, and their depth of penetration into agarose gels are reported.

High density micron-size particles accelerated to high speeds can penetrate deep inside live tissues without inflicting much damage to the cells and have been used to effectively deliver genetic material[1]. This method of ballistic delivery is often called 'biolistics' and the device for shooting particles – 'Gene Gun'. Biolistics proved to be an efficient method for delivery of plasmids and transfection of prokaryotic and eukaryotic (including mammalian) cells [2,3]. Most recently it has been used to stain neuronal tissue with fluorescent dyes[4,5].

A hand-held commercially available version of the gene gun, Helios™ by BioRad (Hercules, CA), uses ~1 $\mu$m in diameter particles made of gold or tungsten, which are accelerated by a short pulse of a high speed flow of Helium (He). Although the carrier particle size has been selected to minimize the cell injury, the high speed jet of He emerging from the gun nozzle can produce severe damage to soft tissue. In order to slow the flow of He, the nozzle expands towards the end, the target is usually shot from a significant distance, and sometimes a mesh filter is placed between the nozzle and the target[5]. Although there is little data on the actual speeds of the particles launched from the Helios gene gun, all those measures result in deceleration of the particles and

---
[a)] Author to whom correspondence should be addressed; electronic mail: agroisman@ucsd.edu

reduction of their penetration depth. A major increase in the depth of penetration into an agarose gel (used to emulate a brain tissue) was reported with a more focused jet of He and short shooting distance, but there was a concomitant increase in damage of the gel surface[6]. A large penetration depth with minimal damage would be very beneficial for staining and transfection of a variety of live tissues, in particular mammalian brain where most of the cell bodies lie 100 $\mu$m below the surface.

In this paper, we propose an alternative design of a pneumatic gun for ballistic particle delivery. Its key feature is the use of active vacuum suction to completely divert the high speed flow of the gas (He) from the gun nozzle, so that there is no damage to the target from the gas flow (Fig.1). The particles are accelerated by continuous high speed flow of He in the inner capillary tube (ICT). The suction is continuously applied through the outer capillary tube (OCT). The diversion occurs over a distance of less than 1 mm, on a time scale of ~2 $\mu$s and is expected to have minimal effect on motion of the particles, which have more inertia than the He gas.

ICT is made of polyamide coated fused silica (MicroFil™ Gauge 23, WPI Inc., Sarasota FL), has an inner diameter $D_i$ = 530 $\mu$m and outer diameter 665 $\mu$m, and is glued to a plastic luer adapter (Fig.1a). OCT is a piece of stainless steel tube with inner and outer diameters of 1.70 and 2.11 mm, respectively. It is made concentric with ICT by two 150 $\mu$m thick "gears" (Fig.1b). The end of OCT is covered by a Ø2.2 mm circular cap with a Ø150 $\mu$m opening in the middle. The cap is 50 $\mu$m thick and has three 100 $\mu$m tall pillars for mechanical strength and self-centering with OCT, to which it is glued (Fig.1b). The cap and the gears are micro-machined out of UV-curable epoxy SU8 (MicroChem, Newton MA) using contact lithography[7]. The opening in the cap, ICT and OCT are coaxial within ~25 $\mu$m. OCT is connected to vacuum through an ~1.5 mm opening in its side at ~20 mm from the cap. We empirically found that the best results are achieved, when the distance between the cap and the end of ICT is in the range of 600-900 $\mu$m.

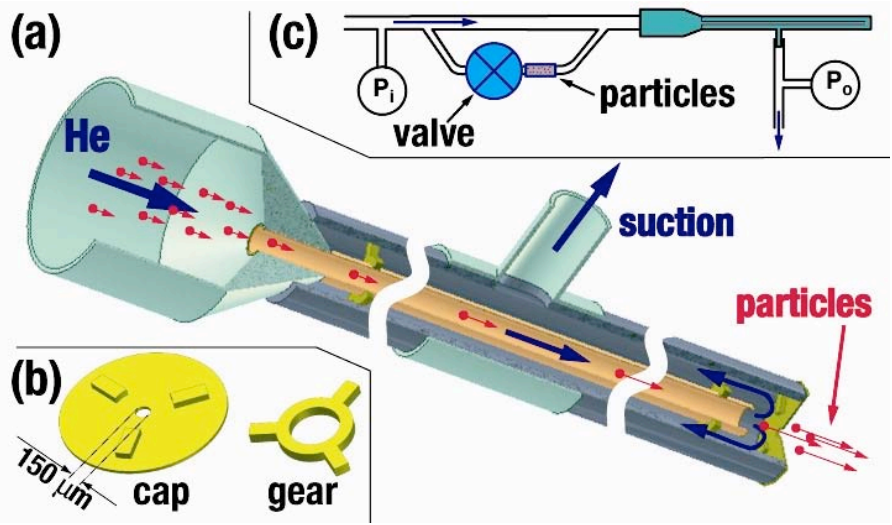

**Fig.1** Schematic drawings showing: (a) design and operation of the capillary gun; (b) micro-machined concentric gear and end cap; (c) connection of the gun to pressurized He and vacuum system.

Four individual guns with distances in that range were tested, and their performance was virtually indistinguishable. The lengths of ICT, $L_i$, were 50-55 mm.

The gun was mounted vertically with the cap facing down and was connected to a vacuum system with a gauge pressure $P_o = -86$ kPa (Fig.1c). The ICT inlet was connected to an adjustable pressure source of compressed He. The He pressure was set at $P_i \approx 120$ kPa, 2-3% below the point where a directed stream of He emerging from the cap could be detected by appearance of ripples on the surface of water ~1 mm under the cap. In order to estimate the average speed of flow at the outlet of ICT, $\bar{v}_o$, we measured volumetric flow rate through the gun, $Q$, with disconnected vacuum suction (by measuring the rate of displacement of water from a graduated cylinder). It was $\bar{v}_o = Q/(\pi D_i^2/4) \cong 630$ m/s. The impact of the flow onto the water surface was less than that of a flow through ICT at 0.2 m/s without the vacuum suction. Therefore, substantially all of the flow of He was diverted to the vacuum system.

In order to inject particles into the flow, the ICT inlet was connected to the compressed He through two parallel lines of thick flexible tubing (Fig.1c). The first line was always open, while the second line was normally kept closed by a solenoid valve. The particles were loaded as a dry powder into a short detachable segment of tubing in the second line downstream from the solenoid valve (Fig.1c). They were "fired" by opening the valve for 300 ms, thus creating a flow of He that introduced the particles into ICT.

To characterize the performance of the gun, we made test shots into agarose gels, which are commonly used to emulate live tissues. We used three types of spherical gold particles (BioRad) that we call A, B and C, with respective diameters, $d$, of 0.47±.15, 1.1±0.1 and 1.27±0.27 µm. The size distributions in the particle samples were characterized under an electron microscope. The gels were inspected under dark field illumination with a 50_/0.5 objective. Representative photographs of a 3% gel, which was imaged at various depths upon shooting with B particles from a distance of 1 mm, are shown in Fig.2a-c. The particles near the plane of image are seen as sharp bright dots, while the out-of-focus particles contribute to the bright background. The particles are spread over a circle of ~150 µm in diameter, which closely matches the opening in the cap. For comparison, we shot the same particles with a Helios gun from ~4 cm at an intermediate 175 psi pressure setting (Fig.2d-f). The particles covered a total area ~1.2 cm in diameter.

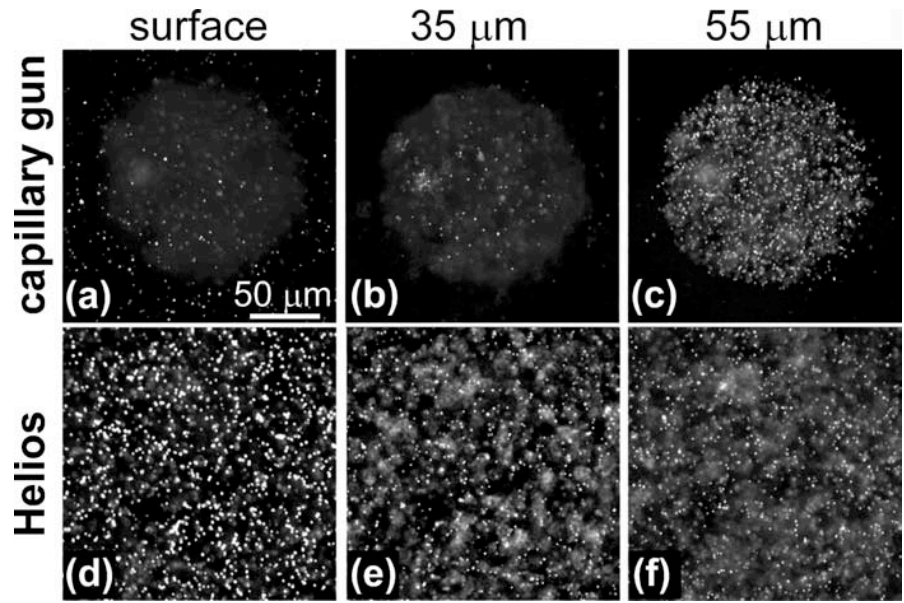

**Fig.2** Dark field microphotographs of a 3% agarose gel with particles B fired from the capillary gun, (a)-(c), and Helios gun at 175 psi, (d)-(f). The image planes are at the gel surface, (a) and (d), at depth 35 μm into the gel, (b) and (e), and 55 μm into the gel, (c) and (f).

It is apparent from the images in Fig.2 that unlike the Helios gun, the capillary gun delivers very few particles to the surface and a notably larger number of particles to 55 µm than to 35 µm depth. The number of particles at different depths, $z$, was counted by taking a stack of images with a step of 3 µm in depth and using Image-Pro software (Media Cybernetics, Silver Spring, MD) to identify bright dots. The distributions of the three kinds of particles by $z$ for shots with the capillary gun and Helios gun at 175 and 120 psi are shown in Fig.3a-c. (The latter pressure is be preferable for staining delicate brain tissue[5].) Each curve represents statistics on ~$10^4$ individual particles. With the capillary gun, the mean $z$ for particles A, B and C are 20, 39 and 38 µm, respectively.

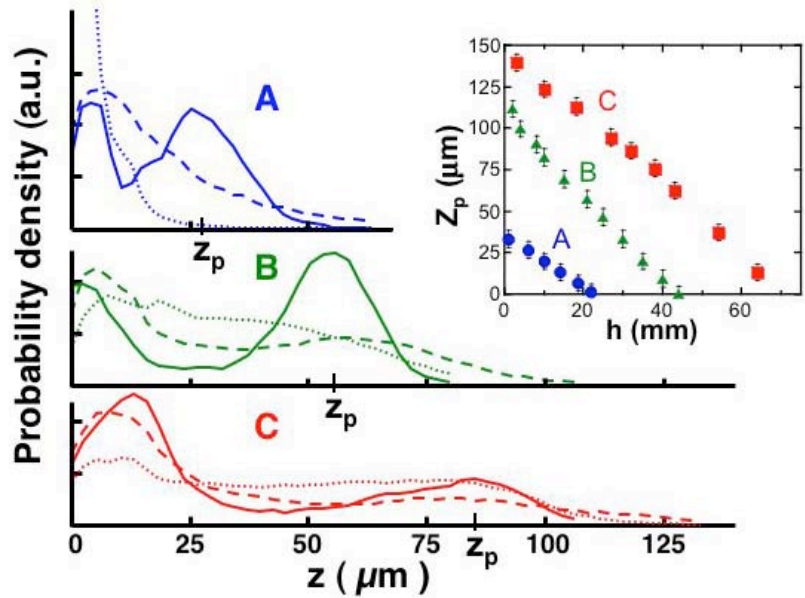

**Fig.3** Distributions of penetration depths, $z$, of particles A, B and C fired into a 3% gel. The solid line is for the capillary gun; positions of the peaks at large $z$ are marked as $z_p$. The dashed and dotted lines are for Helios gun at pressures of 175 and 120 psi, respectively. **Inset:** Characteristic penetration depth into a 0.25% gel, $z_p$, as a function of shooting distance, $h$, in atmosphere of $H_2$ for particles A (circles), B (triangles) and C (squares).

With the Helios Gene Gun, the mean $z$ are 17, 36 and 39 μm and 4.5, 32 and 48 μm at 175 psi and 120 psi, respectively. Thus, penetration depths with the capillary gun are consistently larger for small particles, especially when it is important to minimize the impact of the gas jet on the target. Further, while the depth distributions for the Helios gun are very broad, the distributions for the capillary gun have characteristic peaks at depths $z_p$ near the maximal penetration depths (Fig.3). The peak is narrowest for the most monodisperse sample B, where about 60% of the particles are found at $z$ between 40 and 80 μm.

The particles were ballistically delivered using the capillary gun into 0.25 - 3% agarose gels with no apparent damage to the gel surface other than particle perforation. As expected, mean $z$ decreased with both shooting distance and agarose concentration, due to viscous friction in air and higher resistance to particle motion in denser gels. However, in all cases the distributions of $z$ had the distinct peak near the maximal penetration depth (cf. Fig.3), and we assumed it to be a counterpart of a peak in velocity distribution of particles reaching the gel surface.

To estimate the velocity at that peak, $u_p$, we made shots into a 0.25% gel (where the depth of penetration was maximal) in an atmosphere of Hydrogen ($H_2$) from various distances, $h$, between the cap and the gel surface, and plotted $z_p$ as a function of $h$ (inset in Fig.3). We chose $H_2$ because of its high speed of sound, $v_s \approx 1300$ m/s, and high kinematic viscosity, $\eta_H / \rho_H \approx 1.1 \cdot 10^{-4}$ m²/s, [8] which reduce non-linearity in particle flow resistance associated with finite Mach number, $M = u/v_s$, and Reynolds number, $Re = u d \rho_H / \eta_H$, respectively. (Here $u$ is the particle velocity). In addition, the viscosity of $H_2$, $\eta_H = 9 \cdot 10^{-6}$ Pa·s, [8] is about half the viscosities of air and He. That expands the range of $h$ and improves resolution of the measurements. The dependencies of $z_p$ on $h$ are close to a linear decay for all three kinds of particles (inset in Fig.3).

The condition $z_p = 0$ is met at distances $h_0 = 22$, 44 and 70 mm for particles A, B and C, respectively (inset in Fig.3). We can estimate velocity at the peak of the distribution for particles emerging from the gun, $u_0$, if we assume that $z_p = 0$ corresponds to $u_p = 0$. If we neglect a small correction due to finite $M$, the flow resistance force experienced by the particles can be estimated as[9] $F = -3\pi k d \eta_H u [1 + 0.15 (k \operatorname{Re})^{0.69}]$. Here $k = (1 + 4.5 Kn)^{-1}$ is a correction factor to the Stokes resistance due to finite Knudsen number, $Kn = \lambda / d$, where $\lambda = 0.125$ μm is the mean free path of the $H_2$ molecules[8]. We use the equation of motion, $F = m\dot{u} = \frac{\pi}{6} d^3 \rho_g \dot{u}$, (where $\rho_g = 1.93 \cdot 10^4$ kg/m³ is density of gold) to obtain a differential equation for $u$, and integrate it numerically to obtain $h_0$ as a function of $u_0$ for various $d$. The values of $u_0$ calculated this way for $h_0$ and mean $d$ of particles A, B and C are 400, 230 and 280 m/s, respectively (with estimated errors of about 15%).

It is instructive to compare the speeds of the particles with characteristic speeds of flow in ICT. The average flow speed at the ICT outlet is $\bar{v}_o \approx 630$ m/s, but it is significantly lower upstream from the outlet because of compressibility of He. At the ICT inlet, where the absolute pressure is about 2.2 atm, the average speed of He is estimated as $\bar{v}_i = \bar{v}_o / 2.2 \approx 290$ m/s. Thus, although the speeds of the particles are significantly lower than $\bar{v}_o$, they are comparable with $\bar{v}_i$. This result appears reasonable in view of large characteristic length, $L_a$, required for acceleration of the particles to the high speeds of flow in ICT. The length can be estimated as $L_a \cong \dfrac{v^2}{F/m} \cong \dfrac{\rho_g v d^2}{18 k \eta_{He}[1 + 0.15(k\mathrm{Re})^{0.69}]}$. (Here $Re = v d \rho_{He} / \eta_{He}$ with $\rho_{He} = 0.165$ kg/m$^3$ and $\eta_{He} = 2 \cdot 10^{-5}$ Pa·s, and $k$ is calculated with $\lambda = 0.2$ µm for He molecules[8].) With $v = \bar{v}_o$, $L_a$ for particles B and C is 55 and 67 mm, respectively, which is comparable to the length of ICT, $L_i$. For particles A we obtain a relatively short $L_a$ of about 19 mm, which is a probable reason for their higher characteristic speed.

Increasing $L_i$, while keeping $\bar{v}_o$ the same, requires higher driving pressures, which imply proportionally lower $\bar{v}_i$. Thus, guns with longer ICT that we tested did not give an appreciable increase in the particle penetration depths. However, the depths significantly increased when $D_i$ was expanded from 250 to 530 µm, allowing lower driving pressure. The expansion of ICT might also have reduced negative effects of uncontrolled transverse motion of the particles and inelastic collisions with the walls. Those collisions may be a cause of the wide distributions of the particle penetration depths (Fig.3). Further expansion of ICT proved impractical, though, since it caused a reduction of the velocity of the fastest flow which could be diverted to OCT.

The proposed capillary gun is an easily fabricated alternative to Helios gun. In a pilot experiment we shot B particles coated with GFP expressing reporter plasmid (gWIZ; Aldevron, Fargo, ND) into 293T/17 cells obtained from American Type Culture Collection. After 4 hours fluorescence from expressed GFP was observed in a number of cells. The capillary gun gives large penetration depths for small particles without damaging the surfaces of even the most delicate targets (0.25% agarose gels). It selectively targets small areas and can be inserted into openings down to 2.5 mm in size. So, it may find applications in medicine and live animal biology. A large fraction of the particles is delivered to a narrow interval of depths, and the characteristic penetration depth is reliably controlled by tailoring the shooting distance. The possibility to estimate and control the speed of the particles makes the gun a promising tool to study microscopic mechanical properties of soft materials.

We are very grateful to David Schultz and Jim Glass from Seashell Co. for multiple useful discussions, encouragement and help with the measurements.